\documentclass[prl,twocolumn,showpacs,preprintnumbers,amsmath,amssymb,natbib]{revtex4-1}
\usepackage[dvips]{graphicx}
\usepackage{srcltx}
\usepackage{color}
\usepackage{bm}
\usepackage{amsmath}
\usepackage{amsfonts}
\usepackage{amssymb}
\usepackage{mathrsfs}
\usepackage{boxedminipage}
\usepackage{framed}
\usepackage{bbm}
\usepackage{natbib}

\def\bra#1{\langle #1|}                     % bra vector
\def\ket#1{|#1\rangle}                      % ket vector
\def\matel#1#2#3{\bra{#1}#2\ket{#3}}        % matrix element
\def\mod2#1{\big|#1\big|^2}                 % modulas squared
\def\-{\!-\!}                               % minus sign with less space
\def\+{\!+\!}                               % plus sign with less space
\def\={\;=\;}                               % equal sign for split mode

\def\rr{\vec{r}}

\def\x{\vec{x}}

\def\e{\vec{e}}
\def\0{\vec{0}}

\def\d{\mathrm{d}}                          % infinitesimal
\def\intd#1{\int\d#1\;}                     % 1-dimensional integral
\def\intd2#1{\int\d^2#1\;}                  % 2-dimensional integral
\def\intd3#1{\int\d^3#1\;}                  % 3-dimensional integral
\def\grad{\boldsymbol{\nabla}}              % gradient operator
\def\vec#1{\mathbf{#1}}                     % vector
\def\op#1{\hat{#1}}                         % operator
\def\-{\!-\!}                               % minus sign with less space
\def\+{\!+\!}                               % plus sign with less space
\def\beq{\begin{equation}}                  % begin equation
\def\eeq{\end{equation}}                    % end equation

\def\Moller{M{\o}ller}

\begin{document}

\title{Localization of high-energy electron scattering from atomic vibrations}

\author{C. Dwyer}
\email{c.dwyer@fz-juelich.de}
\affiliation{Ernst Ruska-Centre for Microscopy and Spectroscopy with Electrons, and Peter Gr\"unberg Institute, Forschungszentrum J\"ulich, D-52425 J\"ulich, Germany}

\begin{abstract}%600 characters max
%                   1                     2                     3                     4                     5                     6                     7                     8                     9
%23456789012345678901234567890123456789012345678901234567890123456789012345678901234567890

Electrons with kinetic energies $\sim100$~keV are capable of exciting atomic vibrational states from a distance of microns. Despite such a large interaction distance, our detailed calculations show that the scattering physics permits a high-energy electron beam to locate vibrational excitations with atomic-scale spatial resolution. Pursuits to realize this capability experimentally could potentially benefit numerous fields across the physical sciences.
\end{abstract}


% 61.05.jd Theories of electron diffraction and scattering
% 61.85.+p Channeling phenomena (blocking, energy loss, etc.)
% 68.37.Ma Scanning transmission electron microscopy (STEM)
% 79.20.Uv Electron energy loss spectroscopy

\maketitle

\emph{Introduction.---}Substantial interest surrounds the question of whether it is possible, from the perspective of scattering physics, to use high-energy electrons to locate atomic vibrations with a spatial resolution at, or near, the atomic scale. This interest is triggered by recent reports that it may soon be possible to use an atomic-sized beam in a scanning transmission electron microscope (STEM) to perform spectroscopy at energy resolutions of the order 10 meV \cite{Krivanek2013}, enabling access to vibrational excitations in a transmission geometry. The ability to detect atomic vibrations with high spatial resolution would offer substantial advantages in a number of technologically-important fields, such as catalysis. However, its feasibility has been questioned due to the large degree of ``inelastic delocalization", which measures the distance from which a passing electron can induce an excitation \cite{RitchieHowie1988, *Muller1995}. For a 100~keV electron (typical for a STEM) and an energy loss of 10-100~meV, the delocalization length is estimated to be 1-10~$\mu$m, i.e.\ \emph{far} greater than the size of an atom or molecule, which would imply that high spatial resolution is impossible. 

In this Letter, we demonstrate that, in fact, atomic spatial resolution of vibrational excitations \emph{is} permitted by the scattering physics. We demonstrate this via explicit calculations of vibrational-loss STEM images of selected molecules based on a quantum theory of inelastic electron scattering. We show that, while delocalization effects can be significant, they do not necessarily preclude atomic spatial resolution. The interpretation of the image contrast, however, can be non-trivial. These results will be of central importance to the development of high spatial resolution vibrational spectroscopy as an analytical technique in the physical sciences.

\emph{Background.---}Over the last decade, high spatial resolution electron energy-loss spectroscopy (EELS) in the STEM has become an extremely powerful tool for the analysis of materials. In this technique, an energy-loss spectrum is acquired for each position of the electron beam. The signal from inelastic scattering processes of interest is extracted from each spectrum (by subtracting any other ``background" signals) and plotted as a function of beam position to form an image. In the case of core-level excitations, this technique enables mapping of a material's chemical composition and electronic bonding at the atomic scale \cite{Bosman2007, *Kimoto2007, *Muller2008}, which is ideal for studying interfaces and nanomaterials, for example. In terms of the scattering physics, this spatial resolution is permitted firstly by the atomic-sized beam, and secondly by the inelastic delocalization length $v/\omega$ ($v$ is the electron's velocity and $\hbar\omega$ is the energy loss), which is typically a few \AA ngstroms or less for core-level excitations.

On the other hand, using incident electrons to access vibrational excitations presently requires techniques such as high-resolution EELS (HREELS), which uses low incident energies (a few eV) in a reflection geometry \cite{IbachMills1982, Chesters2006}. The superior energy resolution of this technique can reveal vibrational excitations as sharp peaks in the spectra, at energy losses corresponding to vibrational transitions. HREELS has proven to be extremely powerful in studies of adsorbate molecules on surfaces, for example, where it has enabled fundamental advances in our understanding of molecule-surface interactions. However, the spatial resolution of HREELS is limited, falling well short of the atomic scale. 

\begin{figure*}
\scalebox{0.98}{\includegraphics{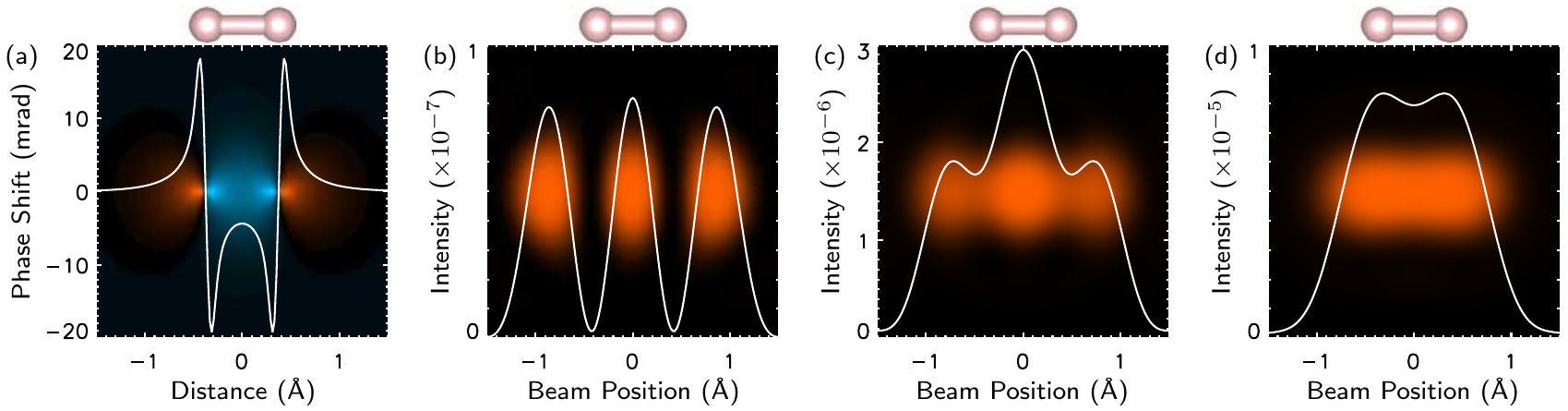}}%
\caption{\label{fig:H2} Calculated vibrational EELS images of an H$_2$ molecule with its molecular axis lying perpendicular to the electron beam: (a) \Moller\ potential for excitation of the H$_2$ stretching mode; (b--d) vibrational EELS images for detector semi-angles 5, 30 and 80 mrad, respectively. Each subfigure shows a square area of size indicated by abscissa values with H$_2$ molecule at centre, while graphs show line traces along the molecular axis. Units in (a) correspond to the phase shift that would be experienced by a 100 keV incident plane wave, and the square of this phase is a measure of scattering probability. Images (b--d) assume an aberation-free 100~keV beam with a 0.7~\AA\ crossover, and the intensity is normalized with respect to the incident beam intensity.}
\end{figure*}

The benefits of combining atomic-spatial and vibrational-energy resolutions could potentially enable fundamental advances in numerous fields. To assess the spatial resolution afforded by the physics, we have used a quantum theory of inelastic electron scattering to calculate vibrational EELS images of selected molecules. We assume a STEM-EELS geometry, whereby the incident electrons form a focused coherent beam, and the images are assumed to be generated by extracting the vibrational signals in analogy with the description above. A 100~keV beam with a convergence semi-angle of 30~mrad is assumed, as appropriate for a state-of-the-art STEM equipped with aberration-corrected beam-forming optics. Such a beam is capable of 0.7~\AA\ spatial resolution. However, in addition to the scattering mechanism itself, the actual spatial resolution achieved is also influenced by the spectrometer's collection angle \cite{Kohl1985}, and so we have considered three collection angles in what follows. Our consideration of the collection angle also has practical implications, in that experimentally this angle influences the spectrometer's energy resolution.

As targets we consider molecular H$_2$ and CO. The simplicity and small size of these molecules allows us to discuss spatial resolution in a straightforward context, and they are exemplary of molecules with and without an electric dipole moment, which has implications that will be made clear below. Moreover, the conclusions we draw will have relevance to potential applications of vibrational EELS imaging in catalysis. 

We employ a theory of molecular vibrations based on quantum mechanics and the harmonic approximation. We use the molecular properties as predicted by density functional theory under the pseudopotential and generalized-gradient approximations \cite{KresseFurthmuller1996, *KresseJoubert1999}. Molecular vibrational properties were computed using the finite-displacement method \cite{Parlinski-etal1997, *Togo-etal2008}, assuming a temperature of 300$^\circ$K.
%In this theory, the atomic displacement of the $\kappa$th atom can be written in the form
%%
%\beq u_\kappa = \left(\frac{\hbar}{2m_\kappa}\right)^{1/2}\sum_{\nu}   \frac{e_{\nu}(\kappa)}{\omega^{1/2}_{\nu}}\left(a_{\nu}+\conj{a}_{\nu}\right),\eeq
%%
%where $m_\kappa$ is the atomic mass, $\nu$ labels the mode of vibration, $e_{\nu}(\kappa)$ is a polarization eigenvector, $\omega_\nu$ is the angular frequency, and $\conj{a}_{\nu}$ and $a_{\nu}$ are the creation and destruction operators, respectively. 

A key quantity in our discussion is the two-dimensional \Moller\ potential for creating one additional quantum in vibrational mode $\nu$, given by 
\beq\begin{split}\label{eq: Moller} V_\nu(x,y) =  \int_{-\infty}^\infty\d z\, \matel{n_\nu + 1}{\op V(\rr)}{n_\nu}e^{-i(\omega_\nu/v) z}\\
= \left(\frac{\hbar(n_\nu+1)}{2\omega_\nu}\right)^{1/2}  \sum_{\kappa} \frac{\e_{\nu}(\kappa)}{m^{1/2}_{\kappa}}\cdot \grad_\kappa V(x,y),\end{split}\eeq
where $\ket{n_\nu}$ is a vibrational state containing $n_\nu$ quanta in mode $\nu$, $\op V(\rr)$ is the Coulomb interaction energy for an electron at position $\rr$, $\hbar\omega_\nu$ is the energy loss, the $z$ axis coincides with the beam direction, $\kappa$ labels the atoms, $\e_{\nu}(\kappa)$ is a polarization vector, $m_\kappa$ an atomic mass, and $\grad_\kappa V(x,y)$ is the gradient of the projected electrostatic potential with respect to the equilibrium position of the nucleus of atom $\kappa$. Loosely speaking, $V_\nu(x,y)$ is the potential that an incident electron ``sees" when it excites the vibrational mode $\nu$. Below we refer to $V_\nu(x,y)$ simply as ``the \Moller\ potential". %\add{The following explicit expression for $V_\nu(x,y)$ (see Ref.~\cite{Rez1977}) will also be useful for later discussion 
%
%\beq\label{eq:scattering function}V_\nu(x,y)
% = \left(\frac{\hbar(\av{n_\nu}+1)}{2\omega_\nu}\right)^{1/2}  \sum_{\kappa} \frac{\e_{\nu}(\kappa)}{m^{1/2}_{\kappa}}\cdot \grad_\kappa V(x,y), \eeq
%
%where $\av{n_\nu}$ is the average population of mode $\nu$ at the given temperature, $\kappa$ labels the atoms, $\e_{\nu}(\kappa)$ is a polarization vector, $m_\kappa$ an atomic mass, and $\grad_\kappa V(x,y)$ is the gradient of the projected electrostatic potential with respect to the equilibrium position of the nucleus of atom $\kappa$.}

The wave function of an inelastically scattered electron which excites mode $\nu$ is given by 
\beq \psi_\nu(x,y;x_0,y_0) = -i\sigma V_\nu(x,y)\psi_0(x-x_0,y-y_0), \eeq
where $\psi_0(x,y)$ is the wave function of the incident electron beam, $(x_0,y_0)$ is the beam position,  and $\sigma$ is an interaction constant \cite{Dwyer2005}. The image intensity obtained for a beam position $(x_0,y_0)$ is given by integrating the inelastic intensity falling within spectrometer's entrance aperture in the far field:
\beq I(x_0,y_0) = \int_{-\infty}^\infty \d u \int_{-\infty}^\infty  \d v\, D(u,v) |\tilde\psi_\nu(u,v;x_0,y_0)|^2, \eeq
where $\tilde\psi_\nu(u,v;x_0,y_0)$ is the Fourier transform of $\psi_\nu(x,y;x_0,y_0)$, and $D(u,v)$ is the detector function which equals unity for positions in the far-field inside the entrance aperture and equals zero otherwise.

\emph{Example 1: H$_2$ molecule.---}We consider an H$_2$ molecule, which we fix fictitiously in free space. This molecule has a bond length calculated to be 0.749~\AA\ (experimental value is 0.750~\AA), and a vibrational stretching mode of calculated energy $\hbar\omega_\mathrm{str.} = 529$~meV (experiment: 517~meV \cite{Dickenson-etal2013}).

\begin{figure*}
\scalebox{0.98}{\includegraphics{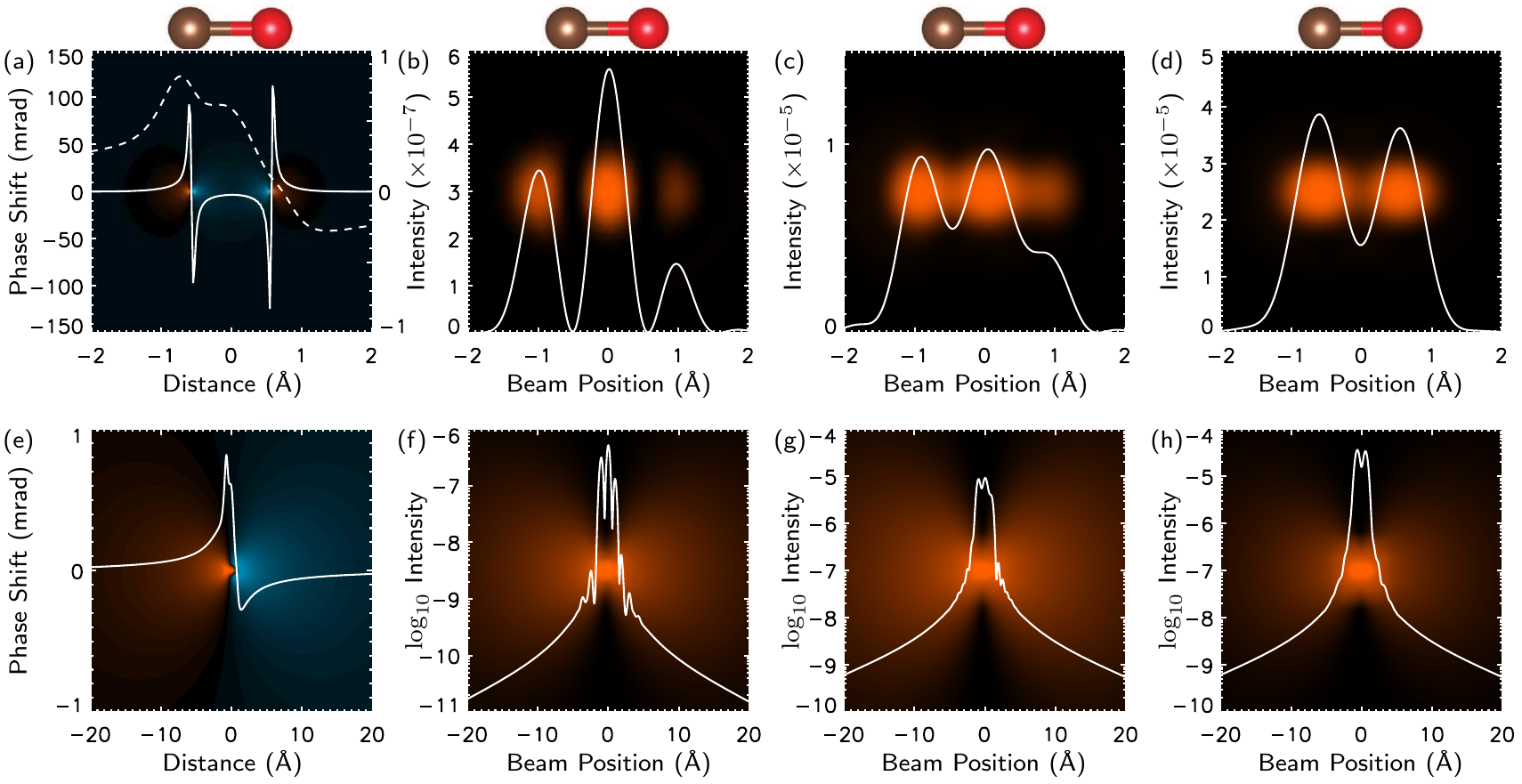}}%
\caption{\label{fig:CO} Effect of dipole scattering on vibrational EELS imaging of a CO molecule lying perpendicular to the electron beam (C on the left): (a) impact potential for excitation of CO stretching mode (solid line, left ordinate) with dipole potential overlaid (dashed line, right ordinate); (b--d) vibrational EELS images for detector semi-angles of 5, 30 and 80 mrad; (e) dipole potential shown over a larger field of view; (f--h) vibrational EELS images shown on a logarithmic scale and over a larger field of view. Beam parameters match Fig.~\ref{fig:H2}.}
\end{figure*}

Fig.~\ref{fig:H2} shows calculated vibrational EELS images of an H$_2$ molecule lying perpendicular to the electron beam \footnote{We use FFT-based multislice calculations with a maximum scattering angle of 316 mrad. The impact potentials (defined later) are derived from the atomic electron scattering factors of Ref.~\cite{Kirkland2010}. Elastic scattering has negligible consequences here and was omitted.}. The \Moller\ potential for excitation of the H$_2$ stretching mode (Fig.~\ref{fig:H2}a) is related to a spatial gradient of the  electrostatic potential (see Eqn.~\eqref{eq: Moller}). As a result, the \Moller\ potential reverses sign at positions which are very close to the equilibrium H positions (where there is a peak in the electrostatic potential). Crucial to our assessment of spatial resolution, we see that, despite a delocalization length of $v/\omega \sim 0.2$~$\mu$m, the \Moller\ potential is contained within an area approximately 2~\AA~$\times$~2~\AA. Therefore we immediately anticipate that the vibrational EELS images (Figs.~\ref{fig:H2}b--d) should be capable of exhibiting atomic spatial resolution, as indeed they do: For a detector semi-angle $\beta=5$~mrad, the vibrational image contains a maximum between the H atoms which, in principle, could be used to locate the position of the H$_2$ molecule with a spatial precision of at least 1 \AA. However, the image intensity is very weak and would be difficult to observe in practice (reducing the convergence angle would increase the normalized intensity at the cost of spatial resolution). Moreover, this vibrational image also contains two strong secondary maxima (three maxima in total), as well as non-intuitive minima close to the equilibrium atomic positions which result from the zeros of the \Moller\ potential.   When $\beta$ is increased to 30~mrad, these minima weaken. For $\beta=80$~mrad the H$_2$ molecule appears as a more-intuitive ``dumbbell" shape with a shallow minimum at the centre. Such behaviour, where atomic-resolution STEM images are easier to interpret if the detector angle is larger than the convergence angle (here 30 mrad), is encountered in other STEM imaging modalities too, for example, bright-field imaging \cite{Xin-etal2012a} and core-level EELS imaging \cite{Oxley2005, *Dwyer2008_doublechanneling}. For $\beta=80$~mrad the peak intensity is comparable to that obtained in core-level EELS from a single oxygen atom using a 100 eV window positioned immediately after the O-$K$ onset.

\emph{Example 2: CO molecule---}The inversion symmetry of the H$_2$ molecule considered in the previous example precludes a very important consideration: A free H$_2$ molecule has no electric dipole moment \footnote{H$_2$ does have an electric quadrupole moment, which is included in Fig.~\ref{fig:H2}, but its affect on spatial resolution is minor so we omit its discussion.}. On the other hand, many molecules possess electric dipole moments arising from the redistribution of charge that takes place during bond formation. Excitation of a vibrational mode will cause the molecule's dipole field to oscillate with period $\omega_\nu$, giving rise to inelastic scattering. Here, the most important point is that the dipole fields are long-ranged, extending over distances much larger than the molecules we are considering. Hence dipole scattering potentially precludes atomic spatial resolution. 

We split the \Moller\ potential into two parts: The first part, called the ``impact potential", arises from changes, due to vibrations, of the molecular charge distribution that would result when electronic bonding is ``switched off". The impact potential is derived from the atomic potentials, and it decays exponentially at large distances so that the scattering from it is inherently localized. The second part of the \Moller\ potential, called the ``dipole potential", arises from changes, due to vibrations, of the so-called ``bonding charge" \footnote{The dipole potentials were calculated by applying the Born-Oppenheimer approximation to all-electron bonding charge density predicted by DFT. By our definition, the dipole potentials also contain the effects of quadrupole and higher-order moments.}.  The names of these potentials  connect with the existing literature on HREELS \cite{IbachMills1982, Thiry-etal1987}. However, while in conventional HREELS is unnecessary to consider the atomic-scale structure of the target when calculating dipole scattering, here it is crucial.

We consider a CO molecule fixed in free space. This molecule has a calculated bond length of 1.142~\AA\ (experiment: 1.128~\AA) and a single stretching mode of calculated energy $\hbar\omega_\mathrm{str.}=265$~meV (experiment: 267 meV \cite{Schonnenbeck-etal1996}). The permanent electric dipole moment was calculated to be 0.143 Debye (experiment: 0.112 Debye \cite{Nelson-etal1967}). While the permanent dipole of CO is relatively small compared to other diatomic molecules, we find that its \emph{dynamic} dipole (due to vibrations) is comparable to that of diatomic molecules with permanent dipoles that are one order of magnitude larger. 

Fig.~\ref{fig:CO} shows the effects of dipole scattering on the vibrational EELS images of a CO molecule lying perpendicular to the electron beam. As in the H$_2$ example above, the impact potential for excitation of the CO stretching mode (Fig.~\ref{fig:CO}a solid line) has zeros at the equilibrium C and O positions. In contrast, the dipole potential (Fig.~\ref{fig:CO}a dashed line, and Fig.~\ref{fig:CO}e) has a large antisymmetric component. It is evident from Fig.~\ref{fig:CO}e that the dipole potential extends far beyond the molecule. However, its amplitude is much smaller than the impact potential. In the vibrational EELS images (Figs.~\ref{fig:CO}b--d), the dipole scattering can introduce considerable asymmetry. For $\beta=5$~mrad, where the dipole scattering makes a significant overall contribution, the effect is such that the CO molecule appears as a ``dumbbell" displaced from its true position. This effect can be interpreted as arising from the quantum interference of impact and dipole scattering. The apparent displacement of the molecule persists for $\beta=30$~mrad. For $\beta=80$~mrad, dipole scattering produces only a minor effect, and now the ``dumbbell" closely coincides with the molecule's true position. Crucially, the vibrational EELS images in Figs.~\ref{fig:CO}b--d exhibit atomic resolution despite a delocalization length of $v/\omega\sim 0.4$~$\mu$m. Figs.~\ref{fig:CO}f--g show the vibrational images on logarithmic scale and over a larger field of view in order to exhibit the ``dipole tails". When the beam is moved 20 \AA\ from the molecule, the image intensity is seen to drop by about 4 orders of magnitude.

\emph{Discussion.---}The CO example demonstrates that, at high spatial resolution, the long-ranged nature of dipole scattering is counterbalanced by its small amplitude: For small detector angles its contribution is comparable to impact scattering, whereas for larger detector angles impact scattering dominates. While we do not claim that this must hold for absolutely all targets, the strength of the CO dynamic dipole is, however, representative of a fairly large class of molecules. In cases where the dipole scattering is even stronger, its effect could be circumvented by employing, for example, an annular detector that excludes the dipole scattering at low angles (though this would also diminish the desirable impact signal). 

Note that this situation is complementary to HREELS, which uses low energy electrons in a reflection geometry, and where dipole scattering is often selected by employing a small acceptance angle \cite{IbachMills1982}. In this context, it is interesting to recall a statement made over 30 years ago by Ibach and Mills: ``...in the impact scattering regime, the total excitation efficiency ($\d S/\d\Omega$) increases as the electron [incident] energy increases. It thus would be most favourable to study large-angle inelastic electron scattering at [incident] energies substantially larger than that used in present generation [low energy] experiments, if suitable spectrometers could be constructed." \cite{IbachMills1982}. Indeed, our detailed calculations show that this idea is very favourable for achieving high spatial resolution.

It was noted earlier that only for a detector angle larger than the convergence angle is the image contrast in Figs.~\ref{fig:H2} and \ref{fig:CO} intuitively interpretable in terms of the molecule's structure. We have also confirmed this behaviour for other atomic structures, including solids and adsorbate molecules on surfaces (to be published elsewhere). In addition to better interpretability, a larger detector has the benefit of a stronger signal. Experimentally, on the other hand, larger detector angles do place greater demands on the spectrometer optics to achieve a given energy resolution, so that a compromise between interpretability and energy resolution is likely to be necessary in practice.

\begin{figure}
\scalebox{1.0}{\includegraphics{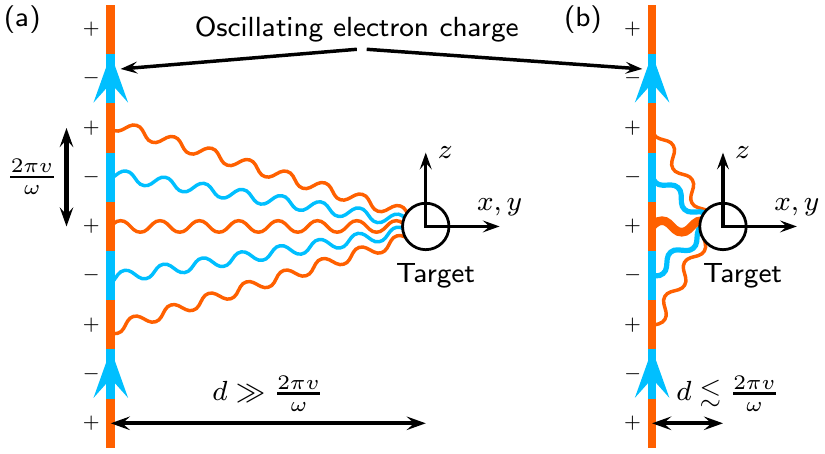}}%
\caption{\label{fig:localization} The physical origin of spatial localization in inelastic electron scattering. (a) If the distance $d$ of closest approach to the target well exceeds the oscillation period $2\pi v/\omega$ of the incident electron's charge distribution then the \Moller\ interaction cancels; (b) If the distance of closest approach is comparable to or smaller than $2\pi v/\omega$ then an interaction is possible.}
\end{figure}

Returning to the general question of delocalization versus spatial resolution in the inelastic scattering of high-energy electrons, considerable insight can be gained by considering the form of \Moller\ potential. This potential represents the interaction between the charge distributions of the scattering electron and the transitioning target. In the case of the electron, the charge distribution incorporates the momentum change along the direction of motion (the consequences of the change in transverse momentum are negligible for high-energy electrons). Hence the electron's charge distribution \emph{oscillates} along the direction of motion with period $2\pi v/\omega$ (Fig.~\ref{fig:localization}). If the electron attempts to interact with the target by passing at a closest distance $d \gg 2\pi v/\omega$, then its oscillatory charge distribution leads to cancellation of the interaction (Fig.~\ref{fig:localization}a). On the other hand, if the closest approach satisfies $d \lesssim 2\pi v/\omega$, then an interaction is possible because portions of positive and negative charge lie at appreciably different distances from the target (Fig.~\ref{fig:localization}b). Mathematically, this leads to an expression for the \Moller\ potential that is exactly analogous to the well-known Yukawa potential describing short-ranged interactions between nucleons. This is accomplished by performing an integration with respect to the $z$ coordinate in Fig.~\ref{fig:localization}, in which case we can obtain
\beq V_\mathrm{\Moller}(\x) = \frac{-e}{2\pi} \int\d^2 \x' \rho(\x')K_0\Bigl(\frac{\omega}{v}|\x-\x'|\Bigr), \eeq
where $\rho$ is the charge distribution of the transitioning target, $K_0$ is a modified Bessel function, bold symbols denote two-dimensional vectors, and $\x$ is the electron's position. The modified Bessel function plays the role of a ``2D Yukawa potential": At small distances it behaves as $\ln|\x|$, while at large distances it behaves as $e^{-(\omega/v)|\x|}/\sqrt{\x}$, that is, exponentially damped, which cuts off the interaction. (As an interesting aside, recalling that in quantum field theory the Yukawa force is ``carried" by particles with mass, here, by analogy, the carrier particles move in two-dimensions with a mass $\hbar\omega/vc$.) 

It should be clear from the present discussion that the quantity $v/\omega$ represents an upper limit on the interaction distance, and that the interaction \emph{can} be strictly more localized than $v/\omega$ if the charge distribution of the target permits. The H$_2$ example above, where there is an absence of a dipole field at large distances, is a case in point. Even when the transitioning target does produce a dipole field at large distances, if this field is sufficiently weak then the interaction can still be effectively more localized than $v/\omega$, as in the CO example above. %In fact, an especially simple analogy exists for elastic scattering, namely, an electron that scatters elastically from a rotationally-symmetric charge-neutral target, such as a closed-shell atom: For elastic scattering there is no cut off since $v/\omega\rightarrow\infty$ as $\omega\rightarrow0$, and yet scattering takes place only within the volume occupied by the atom's charge distribution.

In summary, we have determined that, while the effects of long-ranged dipole scattering and delocalization can be present in inelastic electron scattering from vibrational excitations, the scattering physics permits atomic spatial resolution nonetheless. These results motivate the development of high spatial resolution vibrational spectroscopy as a potentially extremely powerful and unique analytical technique in the physical sciences.

\bibliography{references}

\begin{thebibliography}{24}%
\makeatletter
\providecommand \@ifxundefined [1]{%
 \@ifx{#1\undefined}
}%
\providecommand \@ifnum [1]{%
 \ifnum #1\expandafter \@firstoftwo
 \else \expandafter \@secondoftwo
 \fi
}%
\providecommand \@ifx [1]{%
 \ifx #1\expandafter \@firstoftwo
 \else \expandafter \@secondoftwo
 \fi
}%
\providecommand \natexlab [1]{#1}%
\providecommand \enquote  [1]{``#1''}%
\providecommand \bibnamefont  [1]{#1}%
\providecommand \bibfnamefont [1]{#1}%
\providecommand \citenamefont [1]{#1}%
\providecommand \href@noop [0]{\@secondoftwo}%
\providecommand \href [0]{\begingroup \@sanitize@url \@href}%
\providecommand \@href[1]{\@@startlink{#1}\@@href}%
\providecommand \@@href[1]{\endgroup#1\@@endlink}%
\providecommand \@sanitize@url [0]{\catcode `\\12\catcode `\$12\catcode
  `\&12\catcode `\#12\catcode `\^12\catcode `\_12\catcode `\%12\relax}%
\providecommand \@@startlink[1]{}%
\providecommand \@@endlink[0]{}%
\providecommand \url  [0]{\begingroup\@sanitize@url \@url }%
\providecommand \@url [1]{\endgroup\@href {#1}{\urlprefix }}%
\providecommand \urlprefix  [0]{URL }%
\providecommand \Eprint [0]{\href }%
\providecommand \doibase [0]{http://dx.doi.org/}%
\providecommand \selectlanguage [0]{\@gobble}%
\providecommand \bibinfo  [0]{\@secondoftwo}%
\providecommand \bibfield  [0]{\@secondoftwo}%
\providecommand \translation [1]{[#1]}%
\providecommand \BibitemOpen [0]{}%
\providecommand \bibitemStop [0]{}%
\providecommand \bibitemNoStop [0]{.\EOS\space}%
\providecommand \EOS [0]{\spacefactor3000\relax}%
\providecommand \BibitemShut  [1]{\csname bibitem#1\endcsname}%
\let\auto@bib@innerbib\@empty
%</preamble>
\bibitem [{\citenamefont {Krivanek}\ \emph {et~al.}(2013)\citenamefont
  {Krivanek}, \citenamefont {Lovejoy}, \citenamefont {Dellby},\ and\
  \citenamefont {Carpenter}}]{Krivanek2013}%
  \BibitemOpen
  \bibfield  {author} {\bibinfo {author} {\bibfnamefont {O.~L.}\ \bibnamefont
  {Krivanek}}, \bibinfo {author} {\bibfnamefont {T.~C.}\ \bibnamefont
  {Lovejoy}}, \bibinfo {author} {\bibfnamefont {N.}~\bibnamefont {Dellby}}, \
  and\ \bibinfo {author} {\bibfnamefont {R.~W.}\ \bibnamefont {Carpenter}},\
  }\href@noop {} {\bibfield  {journal} {\bibinfo  {journal} {Microscopy}\
  }\textbf {\bibinfo {volume} {62}},\ \bibinfo {pages} {3} (\bibinfo {year}
  {2013})}\BibitemShut {NoStop}%
\bibitem [{\citenamefont {Ritchie}\ and\ \citenamefont
  {Howie}(1988)}]{RitchieHowie1988}%
  \BibitemOpen
  \bibfield  {author} {\bibinfo {author} {\bibfnamefont {R.~H.}\ \bibnamefont
  {Ritchie}}\ and\ \bibinfo {author} {\bibfnamefont {A.}~\bibnamefont
  {Howie}},\ }\href@noop {} {\bibfield  {journal} {\bibinfo  {journal} {Phil.
  Mag. A}\ }\textbf {\bibinfo {volume} {58}},\ \bibinfo {pages} {753} (\bibinfo
  {year} {1988})}\BibitemShut {NoStop}%
\bibitem [{\citenamefont {Muller}\ and\ \citenamefont
  {Silcox}(1995)}]{Muller1995}%
  \BibitemOpen
  \bibfield  {author} {\bibinfo {author} {\bibfnamefont {D.~A.}\ \bibnamefont
  {Muller}}\ and\ \bibinfo {author} {\bibfnamefont {J.}~\bibnamefont
  {Silcox}},\ }\href@noop {} {\bibfield  {journal} {\bibinfo  {journal}
  {Ultramicroscopy}\ }\textbf {\bibinfo {volume} {59}},\ \bibinfo {pages} {195}
  (\bibinfo {year} {1995})}\BibitemShut {NoStop}%
\bibitem [{\citenamefont {Bosman}\ \emph {et~al.}(2007)\citenamefont {Bosman},
  \citenamefont {Keast}, \citenamefont {Garc\'\i{}a-Mu\~noz}, \citenamefont
  {D'Alfonso}, \citenamefont {Findlay},\ and\ \citenamefont
  {Allen}}]{Bosman2007}%
  \BibitemOpen
  \bibfield  {author} {\bibinfo {author} {\bibfnamefont {M.}~\bibnamefont
  {Bosman}}, \bibinfo {author} {\bibfnamefont {V.~J.}\ \bibnamefont {Keast}},
  \bibinfo {author} {\bibfnamefont {J.~L.}\ \bibnamefont
  {Garc\'\i{}a-Mu\~noz}}, \bibinfo {author} {\bibfnamefont {A.~J.}\
  \bibnamefont {D'Alfonso}}, \bibinfo {author} {\bibfnamefont {S.~D.}\
  \bibnamefont {Findlay}}, \ and\ \bibinfo {author} {\bibfnamefont {L.~J.}\
  \bibnamefont {Allen}},\ }\href@noop {} {\bibfield  {journal} {\bibinfo
  {journal} {Phys. Rev. Lett.}\ }\textbf {\bibinfo {volume} {99}},\ \bibinfo
  {pages} {086102} (\bibinfo {year} {2007})}\BibitemShut {NoStop}%
\bibitem [{\citenamefont {Kimoto}\ \emph {et~al.}(2007)\citenamefont {Kimoto},
  \citenamefont {Asaka}, \citenamefont {Nagai}, \citenamefont {Saito},
  \citenamefont {Matsui},\ and\ \citenamefont {Ishizuka}}]{Kimoto2007}%
  \BibitemOpen
  \bibfield  {author} {\bibinfo {author} {\bibfnamefont {K.}~\bibnamefont
  {Kimoto}}, \bibinfo {author} {\bibfnamefont {T.}~\bibnamefont {Asaka}},
  \bibinfo {author} {\bibfnamefont {T.}~\bibnamefont {Nagai}}, \bibinfo
  {author} {\bibfnamefont {M.}~\bibnamefont {Saito}}, \bibinfo {author}
  {\bibfnamefont {Y.}~\bibnamefont {Matsui}}, \ and\ \bibinfo {author}
  {\bibfnamefont {K.}~\bibnamefont {Ishizuka}},\ }\href@noop {} {\bibfield
  {journal} {\bibinfo  {journal} {Nature}\ }\textbf {\bibinfo {volume} {450}},\
  \bibinfo {pages} {702} (\bibinfo {year} {2007})}\BibitemShut {NoStop}%
\bibitem [{\citenamefont {Muller}\ \emph {et~al.}(2008)\citenamefont {Muller},
  \citenamefont {{Fitting Kourkoutis}}, \citenamefont {Murfitt}, \citenamefont
  {Song}, \citenamefont {Hwang}, \citenamefont {Silcox}, \citenamefont
  {Dellby},\ and\ \citenamefont {Krivanek}}]{Muller2008}%
  \BibitemOpen
  \bibfield  {author} {\bibinfo {author} {\bibfnamefont {D.~A.}\ \bibnamefont
  {Muller}}, \bibinfo {author} {\bibfnamefont {L.}~\bibnamefont {{Fitting
  Kourkoutis}}}, \bibinfo {author} {\bibfnamefont {M.}~\bibnamefont {Murfitt}},
  \bibinfo {author} {\bibfnamefont {J.~H.}\ \bibnamefont {Song}}, \bibinfo
  {author} {\bibfnamefont {H.~Y.}\ \bibnamefont {Hwang}}, \bibinfo {author}
  {\bibfnamefont {J.}~\bibnamefont {Silcox}}, \bibinfo {author} {\bibfnamefont
  {N.}~\bibnamefont {Dellby}}, \ and\ \bibinfo {author} {\bibfnamefont {O.~L.}\
  \bibnamefont {Krivanek}},\ }\href@noop {} {\bibfield  {journal} {\bibinfo
  {journal} {Science}\ }\textbf {\bibinfo {volume} {319}},\ \bibinfo {pages}
  {1073} (\bibinfo {year} {2008})}\BibitemShut {NoStop}%
\bibitem [{\citenamefont {Ibach}\ and\ \citenamefont
  {Mills}(1982)}]{IbachMills1982}%
  \BibitemOpen
  \bibfield  {author} {\bibinfo {author} {\bibfnamefont {H.}~\bibnamefont
  {Ibach}}\ and\ \bibinfo {author} {\bibfnamefont {D.~L.}\ \bibnamefont
  {Mills}},\ }\href@noop {} {\emph {\bibinfo {title} {Electron energy loss
  spectroscopy and surface vibrations}}}\ (\bibinfo  {publisher} {Academic
  Press},\ \bibinfo {address} {London},\ \bibinfo {year} {1982})\BibitemShut
  {NoStop}%
\bibitem [{\citenamefont {Chesters}(2006)}]{Chesters2006}%
  \BibitemOpen
  \bibfield  {author} {\bibinfo {author} {\bibfnamefont {M.~A.}\ \bibnamefont
  {Chesters}},\ }in\ \href@noop {} {\emph {\bibinfo {booktitle} {Handbook of
  vibrational spectroscopy}}},\ \bibinfo {editor} {edited by\ \bibinfo {editor}
  {\bibfnamefont {J.~M.}\ \bibnamefont {Chalmers}}\ and\ \bibinfo {editor}
  {\bibfnamefont {P.}~\bibnamefont {Griffiths}}}\ (\bibinfo  {publisher} {John
  Wiley and Sons},\ \bibinfo {year} {2006})\ p.\ \bibinfo {pages}
  {830}\BibitemShut {NoStop}%
\bibitem [{\citenamefont {Kohl}\ and\ \citenamefont {Rose}(1985)}]{Kohl1985}%
  \BibitemOpen
  \bibfield  {author} {\bibinfo {author} {\bibfnamefont {H.}~\bibnamefont
  {Kohl}}\ and\ \bibinfo {author} {\bibfnamefont {H.}~\bibnamefont {Rose}},\
  }\href@noop {} {\bibfield  {journal} {\bibinfo  {journal} {Adv. Imag. Electr.
  Phys.}\ }\textbf {\bibinfo {volume} {65}},\ \bibinfo {pages} {173} (\bibinfo
  {year} {1985})}\BibitemShut {NoStop}%
\bibitem [{\citenamefont {Kresse}\ and\ \citenamefont
  {Furthm\"uller}(1996)}]{KresseFurthmuller1996}%
  \BibitemOpen
  \bibfield  {author} {\bibinfo {author} {\bibfnamefont {G.}~\bibnamefont
  {Kresse}}\ and\ \bibinfo {author} {\bibfnamefont {J.}~\bibnamefont
  {Furthm\"uller}},\ }\href@noop {} {\bibfield  {journal} {\bibinfo  {journal}
  {Phys. Rev. B}\ }\textbf {\bibinfo {volume} {54}},\ \bibinfo {pages} {11169}
  (\bibinfo {year} {1996})}\BibitemShut {NoStop}%
\bibitem [{\citenamefont {Kresse}\ and\ \citenamefont
  {Joubert}(1999)}]{KresseJoubert1999}%
  \BibitemOpen
  \bibfield  {author} {\bibinfo {author} {\bibfnamefont {G.}~\bibnamefont
  {Kresse}}\ and\ \bibinfo {author} {\bibfnamefont {D.}~\bibnamefont
  {Joubert}},\ }\href@noop {} {\bibfield  {journal} {\bibinfo  {journal} {Phys.
  Rev. B}\ }\textbf {\bibinfo {volume} {59}},\ \bibinfo {pages} {1758}
  (\bibinfo {year} {1999})}\BibitemShut {NoStop}%
\bibitem [{\citenamefont {Parlinski}\ \emph {et~al.}(1997)\citenamefont
  {Parlinski}, \citenamefont {Li},\ and\ \citenamefont
  {Kawazoe}}]{Parlinski-etal1997}%
  \BibitemOpen
  \bibfield  {author} {\bibinfo {author} {\bibfnamefont {K.}~\bibnamefont
  {Parlinski}}, \bibinfo {author} {\bibfnamefont {Z.~Q.}\ \bibnamefont {Li}}, \
  and\ \bibinfo {author} {\bibfnamefont {Y.}~\bibnamefont {Kawazoe}},\
  }\href@noop {} {\bibfield  {journal} {\bibinfo  {journal} {Phys. Rev. Lett.}\
  }\textbf {\bibinfo {volume} {78}},\ \bibinfo {pages} {4063} (\bibinfo {year}
  {1997})}\BibitemShut {NoStop}%
\bibitem [{\citenamefont {Togo}\ \emph {et~al.}(2008)\citenamefont {Togo},
  \citenamefont {Oba},\ and\ \citenamefont {Tanaka}}]{Togo-etal2008}%
  \BibitemOpen
  \bibfield  {author} {\bibinfo {author} {\bibfnamefont {A.}~\bibnamefont
  {Togo}}, \bibinfo {author} {\bibfnamefont {F.}~\bibnamefont {Oba}}, \ and\
  \bibinfo {author} {\bibfnamefont {I.}~\bibnamefont {Tanaka}},\ }\href@noop {}
  {\bibfield  {journal} {\bibinfo  {journal} {Phys. Rev. B}\ }\textbf {\bibinfo
  {volume} {78}},\ \bibinfo {pages} {134106} (\bibinfo {year}
  {2008})}\BibitemShut {NoStop}%
\bibitem [{\citenamefont {Dwyer}(2005)}]{Dwyer2005}%
  \BibitemOpen
  \bibfield  {author} {\bibinfo {author} {\bibfnamefont {C.}~\bibnamefont
  {Dwyer}},\ }\href@noop {} {\bibfield  {journal} {\bibinfo  {journal}
  {Ultramicroscopy}\ }\textbf {\bibinfo {volume} {104}},\ \bibinfo {pages}
  {141} (\bibinfo {year} {2005})}\BibitemShut {NoStop}%
\bibitem [{\citenamefont {Dickenson}\ \emph {et~al.}(2013)\citenamefont
  {Dickenson}, \citenamefont {Niu}, \citenamefont {Salumbides}, \citenamefont
  {Komasa}, \citenamefont {Eikema}, \citenamefont {Pachucki},\ and\
  \citenamefont {Ubachs}}]{Dickenson-etal2013}%
  \BibitemOpen
  \bibfield  {author} {\bibinfo {author} {\bibfnamefont {G.~D.}\ \bibnamefont
  {Dickenson}}, \bibinfo {author} {\bibfnamefont {M.~L.}\ \bibnamefont {Niu}},
  \bibinfo {author} {\bibfnamefont {E.~J.}\ \bibnamefont {Salumbides}},
  \bibinfo {author} {\bibfnamefont {J.}~\bibnamefont {Komasa}}, \bibinfo
  {author} {\bibfnamefont {K.~S.~E.}\ \bibnamefont {Eikema}}, \bibinfo {author}
  {\bibfnamefont {K.}~\bibnamefont {Pachucki}}, \ and\ \bibinfo {author}
  {\bibfnamefont {W.}~\bibnamefont {Ubachs}},\ }\href@noop {} {\bibfield
  {journal} {\bibinfo  {journal} {Phys. Rev. Lett.}\ }\textbf {\bibinfo
  {volume} {110}},\ \bibinfo {pages} {193601} (\bibinfo {year}
  {2013})}\BibitemShut {NoStop}%
\bibitem [{Note1()}]{Note1}%
  \BibitemOpen
  \bibinfo {note} {We use FFT-based multislice calculations with a maximum
  scattering angle of 316 mrad. The impact potentials (defined later) are
  derived from the atomic electron scattering factors of Ref.~\cite
  {Kirkland2010}. Elastic scattering has negligible consequences here and was
  omitted.}\BibitemShut {Stop}%
\bibitem [{\citenamefont {Xin}\ \emph {et~al.}(2012)\citenamefont {Xin},
  \citenamefont {Zhu},\ and\ \citenamefont {Muller}}]{Xin-etal2012a}%
  \BibitemOpen
  \bibfield  {author} {\bibinfo {author} {\bibfnamefont {H.~L.}\ \bibnamefont
  {Xin}}, \bibinfo {author} {\bibfnamefont {Y.}~\bibnamefont {Zhu}}, \ and\
  \bibinfo {author} {\bibfnamefont {D.~A.}\ \bibnamefont {Muller}},\
  }\href@noop {} {\bibfield  {journal} {\bibinfo  {journal} {Microsc.
  Microanal.}\ } (\bibinfo {year} {2012})}\BibitemShut {NoStop}%
\bibitem [{\citenamefont {Oxley}\ \emph {et~al.}(2005)\citenamefont {Oxley},
  \citenamefont {Cosgriff},\ and\ \citenamefont {Allen}}]{Oxley2005}%
  \BibitemOpen
  \bibfield  {author} {\bibinfo {author} {\bibfnamefont {M.~P.}\ \bibnamefont
  {Oxley}}, \bibinfo {author} {\bibfnamefont {E.~C.}\ \bibnamefont {Cosgriff}},
  \ and\ \bibinfo {author} {\bibfnamefont {L.~J.}\ \bibnamefont {Allen}},\
  }\href@noop {} {\bibfield  {journal} {\bibinfo  {journal} {Phys. Rev. Lett.}\
  }\textbf {\bibinfo {volume} {94}},\ \bibinfo {pages} {203906} (\bibinfo
  {year} {2005})}\BibitemShut {NoStop}%
\bibitem [{\citenamefont {Dwyer}\ \emph {et~al.}(2008)\citenamefont {Dwyer},
  \citenamefont {Findlay},\ and\ \citenamefont
  {Allen}}]{Dwyer2008_doublechanneling}%
  \BibitemOpen
  \bibfield  {author} {\bibinfo {author} {\bibfnamefont {C.}~\bibnamefont
  {Dwyer}}, \bibinfo {author} {\bibfnamefont {S.~D.}\ \bibnamefont {Findlay}},
  \ and\ \bibinfo {author} {\bibfnamefont {L.~J.}\ \bibnamefont {Allen}},\
  }\href@noop {} {\bibfield  {journal} {\bibinfo  {journal} {Phys. Rev. B}\
  }\textbf {\bibinfo {volume} {77}},\ \bibinfo {pages} {184107} (\bibinfo
  {year} {2008})}\BibitemShut {NoStop}%
\bibitem [{Note2()}]{Note2}%
  \BibitemOpen
  \bibinfo {note} {H$_2$ does have an electric quadrupole moment, which is
  included in Fig.~\ref {fig:H2}, but its affect on spatial resolution is minor
  so we omit its discussion.}\BibitemShut {Stop}%
\bibitem [{Note3()}]{Note3}%
  \BibitemOpen
  \bibinfo {note} {The dipole potentials were calculated by applying the
  Born-Oppenheimer approximation to all-electron bonding charge density
  predicted by DFT. By our definition, the dipole potentials also contain the
  effects of quadrupole and higher-order moments.}\BibitemShut {Stop}%
\bibitem [{\citenamefont {Thiry}\ \emph {et~al.}(1987)\citenamefont {Thiry},
  \citenamefont {Liehr}, \citenamefont {Pireaux},\ and\ \citenamefont
  {Caudano}}]{Thiry-etal1987}%
  \BibitemOpen
  \bibfield  {author} {\bibinfo {author} {\bibfnamefont {P.~A.}\ \bibnamefont
  {Thiry}}, \bibinfo {author} {\bibfnamefont {M.}~\bibnamefont {Liehr}},
  \bibinfo {author} {\bibfnamefont {J.~J.}\ \bibnamefont {Pireaux}}, \ and\
  \bibinfo {author} {\bibfnamefont {R.}~\bibnamefont {Caudano}},\ }\href@noop
  {} {\bibfield  {journal} {\bibinfo  {journal} {Physica Scripta}\ }\textbf
  {\bibinfo {volume} {35}},\ \bibinfo {pages} {368} (\bibinfo {year}
  {1987})}\BibitemShut {NoStop}%
\bibitem [{\citenamefont {Sch{\" o}nnenbeck}\ \emph {et~al.}(1996)\citenamefont
  {Sch{\" o}nnenbeck}, \citenamefont {Cappus}, \citenamefont {Klinkmann},
  \citenamefont {Freund}, \citenamefont {Petterson},\ and\ \citenamefont
  {Bagus}}]{Schonnenbeck-etal1996}%
  \BibitemOpen
  \bibfield  {author} {\bibinfo {author} {\bibfnamefont {M.}~\bibnamefont
  {Sch{\" o}nnenbeck}}, \bibinfo {author} {\bibfnamefont {D.}~\bibnamefont
  {Cappus}}, \bibinfo {author} {\bibfnamefont {J.}~\bibnamefont {Klinkmann}},
  \bibinfo {author} {\bibfnamefont {H.~J.}\ \bibnamefont {Freund}}, \bibinfo
  {author} {\bibfnamefont {L.~G.~M.}\ \bibnamefont {Petterson}}, \ and\
  \bibinfo {author} {\bibfnamefont {P.~S.}\ \bibnamefont {Bagus}},\ }\href@noop
  {} {\bibfield  {journal} {\bibinfo  {journal} {Surf. Sci.}\ }\textbf
  {\bibinfo {volume} {347}},\ \bibinfo {pages} {337} (\bibinfo {year}
  {1996})}\BibitemShut {NoStop}%
\bibitem [{\citenamefont {{Nelson Jr.}}\ \emph {et~al.}(1967)\citenamefont
  {{Nelson Jr.}}, \citenamefont {{Lide Jr.}},\ and\ \citenamefont
  {Maryott}}]{Nelson-etal1967}%
  \BibitemOpen
  \bibfield  {author} {\bibinfo {author} {\bibfnamefont {R.~D.}\ \bibnamefont
  {{Nelson Jr.}}}, \bibinfo {author} {\bibfnamefont {D.~R.}\ \bibnamefont
  {{Lide Jr.}}}, \ and\ \bibinfo {author} {\bibfnamefont {A.~A.}\ \bibnamefont
  {Maryott}},\ }\href@noop {} {\emph {\bibinfo {title} {Selected values of
  electric dipole moments for molecules in the gas phase}}},\ \bibinfo {type}
  {Tech. Rep.}\ \bibinfo {number} {10}\ (\bibinfo  {institution} {National
  Bureau of Standards},\ \bibinfo {year} {1967})\BibitemShut {NoStop}%
\end{thebibliography}%

\end{document}